\documentclass[prd,article,nofootinbib,twocolumn,preprintnumbers,superscriptaddress]{revtex4}
\usepackage{graphicx,amsfonts,color,comment,amsmath,hyperref,float}
\usepackage{amssymb}	

\def\aap{\ref@jnl{A\&A}}                

\usepackage{mathrsfs,amssymb}  
\usepackage{cancel}
\usepackage[normalem]{ulem}

\newcommand{\be}{\begin{equation}}
\newcommand{\ee}{\end{equation}}
\newcommand{\bea}{\begin{eqnarray}}
\newcommand{\eea}{\end{eqnarray}}
\begin{document}

\title{Constraints from the time lag between gravitational waves and gamma rays: Implications of GW170817 and GRB 170817A}

\author{Ian M. Shoemaker}
\affiliation{Department of Physics, University of South Dakota, Vermillion, SD 57069, USA}

\author{Kohta Murase}
\affiliation{Department of Physics; Department of Astronomy \& Astrophysics; Center for Particle and Gravitational Astrophysics, The Pennsylvania State University, University Park, PA 16802, USA}
\affiliation{Yukawa Institute for Theoretical Physics, Kyoto University, Kyoto 606-8502, Japan}

\date{\today}
\begin{abstract}
The Laser Interferometer Gravitational-Wave Observatory (LIGO) has recently discovered gravitational waves (GWs) from its first neutron star-neutron star merger at a distance of $\sim 40$~Mpc from the Earth. The associated electromagnetic (EM) detection of the event, including the short gamma-ray burst within $\Delta t \sim 2$~s after the GW arrival, can be used to test various aspects of sources physics and GW propagation. Using GW170817 as the first GW-EM example, we show that this event provides a stringent direct test that GWs travel at the speed of light. The gravitational potential of the Milky Way provides a potential source of Shapiro time delay difference between the arrival of photons and GWs, and we demonstrate that the nearly coincident detection of the GW and EM signals can yield strong limits on anomalous gravitational time delay, through updating the previous limits taking into account details of Milky Way's gravitational potential. Finally, we also obtain an intriguing limit on the size of the prompt emission region of GRB 170817A, and discuss implications for the emission mechanism of short gamma-ray bursts. 
\end{abstract}
\preprint{}


\maketitle

\section{Introduction}
New observations of gravitational phenomena provide an opportunity for new discoveries, or more minimally stringent constraints on theories of gravity. The recent observation of Gravitational Waves (GWs) from Black Hole - Black Hole (BH-BH) mergers by the advanced Laser Interferometer Gravitational-Wave Observatory (advanced-LIGO) and the Virgo interferometer have ushered in a new window on the Universe for astrophysics, but also for fundamental tests of our understanding of gravity~\cite{Abbott:2016blz,Abbott:2017oio}. 
Now the LIGO Collaboration has also observed the first neutron star -- neutron star merger (NS-NS) event, GW170817, with high significance electromagnetic (EM) counterparts at different wavelengths~(e.g.,~\cite{PhysRevLett.119.161101,2041-8205-848-2-L12,2041-8205-848-2-L14,2041-8205-848-2-L15,2041-8205-848-2-L20,Troja:2017nx,Evanseaap9580,2041-8205-848-2-L31,Kasliwal:2017ngb}). The significance of the observed EM counterparts with a GW event is hard to overstate. In addition to representing a enormous milestone for multi-messenger astrophysics, it also gives us an unprecedented test of General Relativity (GR).  {In Ref.~\cite{Monitor:2017mdv}, some implications for both astrophysics and fundamental physics have been considered independently of this work.}

Among various EM counterparts from radio, IR, optical, X-rays, and gamma rays, gamma-ray emission was detected by the {\it Fermi}-GBM and {\it INTEGRAL} detectors. The advanced-LIGO triggered the NS-NS event $\sim1.7$~s prior to the GBM trigger, and the observed duration of gamma-ray emission was $2.0\pm0.5$~s~\cite{2041-8205-848-2-L12,2041-8205-848-2-L14}. The emission is consistent with short gamma-ray bursts (SGRB) with a typical duration of $\sim2$~s, which supports the hypothesis that the progenitors of SGRBs are NS-NS mergers.
One may express the observational time delay as  
\begin{equation}
{\Delta t}_{\rm obs}={\Delta t}_{\rm ast}+{\Delta t}_{\rm non-GR},
\end{equation}
where ${\Delta t}_{\rm ast}$ is the astrophysical time delay caused by the fact that the gamma-ray emission region should be larger than the coalescence site, whereas ${\Delta t}_{\rm non-GR}$ is the time delay caused by possible non-GR effects including the violation of the weak equivalence principle and massive gravitons.  

In light of the GW150914 BH-BH merger event, a number of constraints on modifications to GR were obtained in~\cite{TheLIGOScientific:2016src}. For example, the LIGO collaboration obtained directly one of the strongest model-independent bounds on the graviton mass, $m_{g} \le 1.2 \times 10^{-22}~{\rm eV}$. Other methods include testing the GW speed if the GW signal was strongly lensed~\cite{Fan:2016swi}. The disputed association of GW150914 with the transient source with $> 50$ keV photons observed by the {\it Fermi} Gamma-Ray Burst Monitor (GBM) was similarly used to derive tentative constraints on Lorentz violation in the graviton propagation~\cite{Ellis:2016rrr,Yunes:2016jcc}, the weak equivalence principle~\cite{Wu:2016igi}, certain properties of extra dimensional models~\cite{Yu:2016tar}, and most directly differences from the relation, $c_{\rm gw}=c$. 

Contrary to the previous claims, GW170817 can be regarded as the first convincing example of the event with EM counterparts. 
In this paper we examine in light of this first established GW signal and EM association the constraints one can derive on various non-GR effects as well as on source physics. Given the observed time delay between the GW and EM signals of $\Delta t=1.74\pm0.05$~s~\cite{Monitor:2017mdv} and the observed ``electromagnetic'' distance to NGC 4993 of $D=41\pm3.1$~Mpc ($D_{GW} = 43.8^{+2.9}_{-6.9}$ Mpc)~\cite{2041-8205-848-2-L31}, we can set new stringent limits on the size of the gamma-ray emission region, GW propagation speed, and provide new tests on effects of the gravitational time delay. {In the latter, we will examine the impact of modeling the Milky Way's gravitational potential on these limits, thus extending and refining the initial analysis made in Ref.~\cite{Monitor:2017mdv}. }

\section{Implications for the Gamma-Ray Emission Site} 
First, we consider the astrophysical time delay with ${\Delta t}_{\rm non-GR}=0$ in Eq.~(1). The assumption of ${\Delta t}_{\rm ast}=0$ is not realistic because the emission region is far from of the central object -- a black hole or neutron star. There is the so-called compactness problem, in which gamma rays cannot escape from the too compact region due to $\gamma\gamma\rightarrow e^+e^-$. In the context of gamma-ray bursts~\cite{Meszaros:2006rc}, this can be solved when the source moves relativistically. 
Assuming that the source, which is likely to be a relativistic jet, moves with the velocity $v_j$, and the gamma-ray emission occurs at the radius $r_{\rm em}$, the time delay between GW and gamma rays is given by
\begin{equation}
{\Delta t}_{\rm ast}=\frac{r_{\rm em}}{v_{j}}\left(1-\frac{v_{j}}{c}\cos\theta_{\rm ob}\right),
\end{equation}
where $\theta_{\rm ob}$ is the angle between the jet axis and line of sight to the source. Knowing the size of the emission region is important to understand the physical mechanism of prompt emission~\cite{Murase:2007ya,Kumar:2014upa}.
{ For a top hat jet with a finite opening angle with $\theta_j\sim10^\circ$, we may replace $\theta_{\rm ob}$ with $\theta_{\rm ob}-\theta_j$.}

With { $\theta_{\rm ob}\lesssim\theta_j$}, the time delay be is given by ${\Delta t}_{\rm ast}=r_{\rm em}/(2\gamma_j^2c)$, where $\gamma_j$ is the jet Lorentz factor. While low-luminosity SGRBs cannot be excluded only by the observation of the prompt emission, the possibility of on-axis, highly relativistic jets with $\theta_{\rm ob}\lesssim\theta_j$ is constrained by x-ray and radio observations~\cite{2041-8205-848-2-L20,Evanseaap9580,Kasliwal:2017ngb}. With ${\Delta t}_{\rm ast}\leq2$~s and $\theta_{\rm ob}-\theta_j\sim20^\circ$~\cite{2041-8205-848-2-L20,Troja:2017nx,Evanseaap9580}, which is consistent with the late-time afterglow data, the emission radius is constrained to be
\be
r\lesssim c{\Delta t}_{\rm ast}{(1-\cos\theta_{\rm ob})}^{-1}\simeq9.9\times{10}^{11}~{\rm cm}.
\ee
{ Note that the above constraint is applied if the emission comes from the jet.} 
This upper limit can be compatible with the photospheric radius. With a typical isotropic-equivalent luminosity of $L_{\rm iso}\sim10^{50.5}~{\rm erg}~{\rm s}^{-1}$ and $\gamma_j\sim100$, the photospheric radius is estimated to be $r_{\rm ph}=L_{\rm iso}\sigma_T/(4\pi \gamma_j^3m_pc^3)\simeq3.7\times{10}^{11}$~cm. 
{ However, as long as we consider a top hat jet, the off-axis emission from a highly relativistic jet is highly suppressed for $\theta_{\rm ob}\sim30^\circ$. 
Thus, in order to explain the observed luminosity, $L_{\gamma}\approx1.1\times{10}^{47}~{\rm erg}~{\rm s}^{-1}$, a structured jet with a slow jet component} with a wide opening angle may also be invoked. 
In addition, more realistically, the jet may propagate with a subrelativistic or mildly relativistic speed in the merger ejecta~\cite{Nagakura:2014hza}, and makes a contribution to the time delay. Also, the jet formation has not been clearly seen yet in the latest numerical relativity simulations~\cite{Kiuchi:2009jt,Kiuchi:2017zzg}, and there could be a time offset between the jet launch and GW emission at the coalescence, which can also cause an additional time delay. 

Based on the constraints we obtained, we suggest the emission from the jet-induced breakout emission from the merger ejecta as one of the possible scenarios~(see also~\cite{Lazzati2017,Kasliwal:2017ngb}). Note that the observed duration of the gamma-ray spike is $\delta t\sim0.5$~s, which was found in {\it Fermi}-GBM~\cite{2041-8205-848-2-L14}.
If the effective jet speed in the ejecta is subrelativistic, given that the merger ejecta is launched with a subrelativistic velocity, $V\sim0.3$~c, the emission radius can be $r_{\rm em}\approx V{\Delta t}_{\rm ast}\simeq1.7\times{10}^{10}~(V/0.3c)$~cm, which may lead to emission with a duration of $\sim r_{\rm em}/c\sim0.6$~s. { This radius should be regarded as the minimum radius, and a larger radius is favored to avoid the compactness problem for gamma rays. Also, the outer envelope of the merger ejecta is usually extended to larger radii.}
In more realistic situations, the emission region may be more mildly relativistic with $\gamma_{c}\sim2$, and for a quasi-isotropic outflow we may use the on-axis relationship, ${\Delta t}_{\rm ast}\approx r_{\rm em}/(2\gamma_c^2c)$. Then, the size of the emission region is estimated to be $r_{\rm em}\approx 2\gamma_c^2\Delta t_{\rm ast}\lesssim4.8\times{10}^{11}~{(\gamma_c/2)}^2$~cm, which can be compatible with the observations of gamma rays. The latter case is expected for the cocoon breakout (e.g.,~\cite{Lazzati2017}) or trans-relativistic ejecta that is formed by a choked jet~(e.g.,~\cite{Meszaros:2001ms,Murase:2013ffa}).   

Note that the assumptions that the outflow is launched at the same time of the GW emission and $v_j\approx c$ are rather conservative. If the observed time delay is explained by astrophysical effects, in principle, we could improve the bounds on non-GR effects in the GW propagation, which we discuss below.  

\section{Bounds on the Propagation Speed of Gravitational Waves} 

In GR, GWs propagate with a speed of $c_{\rm gw}=c$, but the time difference can be expected in some modified GR theories. 
It was recently pointed out that even in the absence of an EM counterpart, the speed of GWs can be bounded using the timing information between widely spaced detectors~\cite{Cornish:2017jml}. Using the first three GW detections from BH mergers, they found $0.55 c < c_{\rm gw} < 1.42c$. 

Assuming that GWs and gamma-rays are emitted {\it at the same time}, the time delay between GW and EM signals is written as: 
\begin{equation}
{\Delta t}=\frac{D}{c}\left(1-\frac{c}{c_{\rm gw}}\right).
\end{equation}
Then, we find that the deviation from the EM speed can be deduced to be 
\be 
\left(1- \frac{c_{\rm gw}}{c}\right) = 5 \times 10^{-16} \left(\frac{40~{\rm Mpc}}{D}\right)~\left(\frac{\Delta t}{2~{\rm s}}\right).
\ee

Note that other stringent, albeit {\it indirect}, tests of GW speed can be obtained from the non-observation of gravitational Cherenkov radiation from high-energy cosmic rays~\cite{Moore:2001bv}.  

\vspace{1cm}

\section{Bounds on Weak Graviton Mass} 

A nonzero graviton mass may also contribute to the time dispersion in GW signals. In this case the graviton speed is diminished relative to the massless photon as, $c_{\rm gw}=\sqrt{1- m_{g}^{2}/E^{2}}$, where $E$ is the graviton energy. 
Then the relative time delay between a photon and graviton signal will be 
\be
\Delta t\approx \frac{m_{g}^{2}}{2H_{0}} \left(\frac{1}{E_{1}^{2}} - \frac{1}{E_{2}^{2}}\right)~ \int_{0}^{z_{s}} \frac{(1+z')~dz'}{h(z')}
\ee
where $h_{z} \equiv H(z)/H_{0}$, with $H(z)$ the redshift dependent Hubble parameter and $H_{0} = 68~{\rm km}~{\rm s}^{-1}~{\rm Mpc}^{-1}$ the current expansion rate. The parameter $z_{s}$ is the source redshift.

The existing analysis of the BH-BH merger in \cite{TheLIGOScientific:2016src}, resulted in dramatically strong bounds, $m_{g} \le 1.2 \times 10^{-22}~{\rm eV}$, as a result of the detailed waveform information. We find that GW170817 imposes no stronger bound since the source is so close, but the LIGO collaboration may be able to improve on this limit from a similar waveform analysis as done previously in Ref.~\cite{TheLIGOScientific:2016src}. 

Note that in Ref.~\cite{Bose:1988wi} the authors considered the simultaneous impact of weak equivalence principle with nonvanishing $\Delta \gamma$ and finite graviton mass. 

\begin{figure}[b!]
\begin{center}
\includegraphics[width=.45\textwidth]{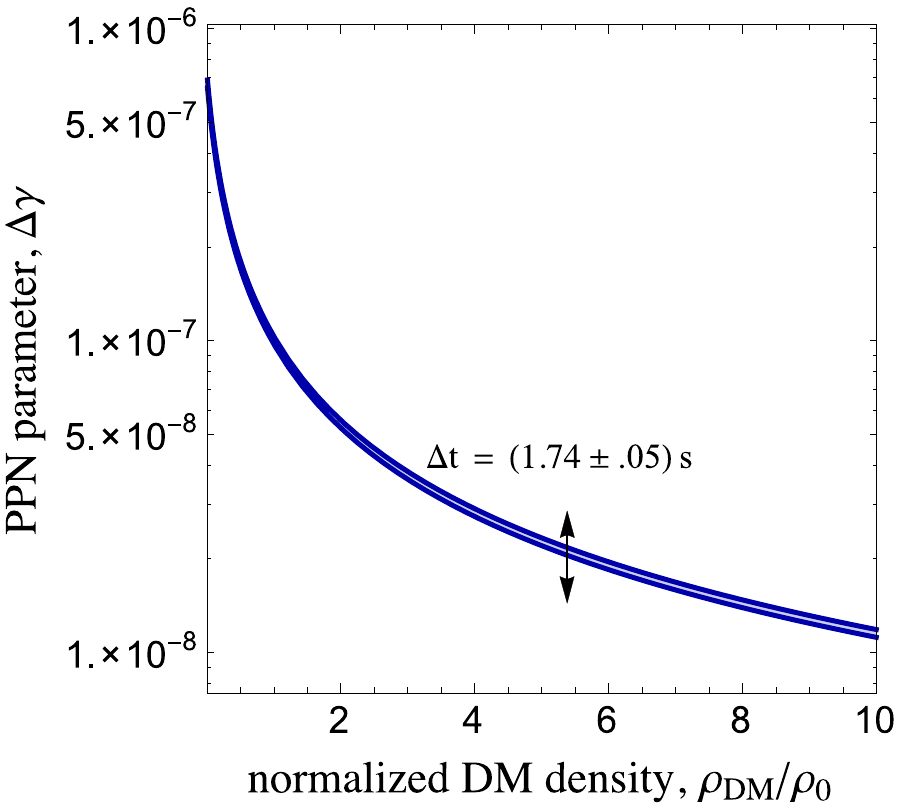} 
\caption{The dependence on the PPN bound as a function of the DM density, normalized to the canonical local value $\rho_{0} \equiv 0.3~{\rm GeV}~{\rm cm}^{-3}$. The shaded blue region shows how the bound changes within the time window uncertainty. 
}
\label{fig1}
\end{center}
\end{figure}

\section{Bounds on Gravitational Time Delay Difference} 
A classic test of GR, the original Shapiro time delay was a proposal to use the ``echoes'' of radar pulses directed toward the inner solar system to test the predicted time delay photons should experience as they sample the gravitational potential~\cite{Shapiro:1964uw}.

A crucial hallmark of Einstein's Theory of GR is the weak equivalence principle. However this principle is an outcome of the metric nature of GR, and thus any metric theory of gravity whether GR or otherwise will similarly predict that test particles will follow the same trajectories. Thus an observed anomalous Shapiro time delay between the arrival of different test particles (photons, neutrinos, gravitons, etc.) can bound violations of the weak equivalence principle.  {The use of time delays between various messenger particles/waves to place constraints on the Equivalence Principle has a long history~(e.g.~\cite{Krauss:1987me,bull,Desai:2008vj}).}

To derive the impact of this effect we assume that any observed time delay is dominantly controlled by the Shapiro time delay due to the Milky Way gravitational potential. Similar constraints on the weak equivalence principle for neutrinos and photons were obtained in the aftermath of the SN1987A event~\cite{Longo:1987gc,Krauss:1987me}. Related constraints were obtained on long-range neutrino interactions as well~\cite{Pakvasa:1988gd}. To date the strongest limits on weak equivalence principle tests come from FRBs, limiting $\Delta \gamma \lesssim 4.4 \times 10^{-9}$~\cite{Wei:2015hwd}. 

Using the standard notation in the post-newtonian approximation, metric theories of gravity have spatial and temporal metric components written as
\bea g_{00} &=& -\left[ 1 - 2 U(r)  \right] \\
 g_{ij}&=& \delta_{ij} \left[ 1 + 2 \gamma U(r) \right]dx_{i} dx^{j}
\eea
where $\gamma$ is one of the ten parameterized post-Newtonian (PPN) parameters. The parameter $\gamma$ is a measure of the spatial curvature experienced by test particles. Then one can compute the coordinate time delay as
\be 
\tau=  \int \sqrt{\frac{g_{{\rm xx}}}{-g_{00}}} d{\rm x}
\ee

Assuming a Keplerian form for the potential $U(r) =-GM/r$ and keeping only the part proportional to $\Delta \gamma$ one can estimate the time delay as~\cite{Misner:1974qy,Longo:1987gc}
%
\begin{eqnarray}
\Delta t&=&\Delta\gamma ~GM_{{\rm MW}}\nonumber\\
&\times&\ln \left\{\frac{ \left[d+ (d^{2} -b^{2})^{1/2}\right]\left[r_{G} +s_{n}(r_{G}^{2} -b^{2} )^{1/2}\right]}{b}^{2} \right\}, \,\,\,\,\,\,\,\,
\end{eqnarray}
where $r_{G}= 8.3$ kpc is the galactic center, $b$ is the impact parameter, and $d$ is the distance to the source. The discrete parameter $s_{n} = \pm 1$ is a correction taking the value $+1$ for sources along the direction of the Milky Way and $-1$ for sources pointing away~\cite{Wu:2016igi}. The Milky Way mass is roughly $M_{MW} \approx 5 \times 10^{11}~M_{\odot}$~\cite{McMillan:2011wd,Kafle:2012az}. 

The impact parameter can be found from the source direction, 
\bea 
b &=& r_{G} \sqrt{1- (\sin \delta_{S} \sin \delta_{G} + \cos \delta_{S} \cos \delta_{G}  \cos (\beta_{S} - \beta_{G}))^{2}}  \nonumber \\ 
 &&
\eea
using the source and galactic center right ascension ($\beta$) and declination ($\delta$). 
%

Given the observed time delay of $\Delta t \approx 2$~s and the distance to NGC 4993 of $\approx 40$ Mpc we can set the limit
\be 
\Delta \gamma |_{{\rm Keplerian}} \lesssim 7.4 \times 10^{-8}
\ee
This estimate is meant only as an illustration of the order of magnitude for the constraint. Let us now investigate the constraint on the gravitational time delay difference in a more complete description of the Milky Way's potential.

The above treatment of the Milky Way as a point source can be improved as follows. Extending the analysis to a more realistic potential comprised of bulge, disk and dark matter (DM) halo components can be done straightforwardly. Here we model the disk as a spherical potential of the form
\be
U_{d} (r) =  -\frac{GM_{d}}{r} \left(1- e^{-r/r_{d}}\right),
\ee
where the disk mass is {$M_{d} = (5.17 \pm 1.11) \times 10^{10}~M_{\odot}$}~\cite{Licquia:2014rsa}, and $r_{d} = (2.6\pm0.5)$ kpc. 

For the bulge component a simple Keplerian treatment is adequate 
\be 
U_{b}(r) =- \frac{G M_{b}}{r},
\ee
where the bulge mass is taken to be { $M_{b} = (2\pm0.3) \times 10^{10}~M_{\odot}$}~\cite{2016A&amp;A...587L...6V}. The DM halo component can be found by solving the Poisson equation with a NFW density profile
\be 
\rho(r)_{{\rm NFW}} = \frac{\rho_{s}}{(r/r_{s})(1+ (r/r_{s}))^{2}}
\ee
{where $r_{s} = 16.1^{+17}_{-7.8}$ kpc and $\rho_{s} = 1.4^{+2.9}_{-0.93} \times 10^{7}~M_{\odot}~{\rm kpc}^{-3}$~\cite{Nesti:2013uwa}.} 

Taking into account the above uncertainties on the MW density profile we find that this more detailed model results in the bound on the PPN parameter at 90$\%$ CL, 
\be 
\Delta \gamma |_{{\rm MW~model}} \lesssim 8.3 \times 10^{-8}.
\ee
{with the dominant source of modeling error coming from the scale radius of the NFW dark matter profile.}

We see that the more realistic treatment of Milky Way's gravitational potential results in a weakening of the derived bound. This is expected however since the DM halo is significantly more disperse than a point mass at the Galactic Center. In addition we can scan more broadly the dependence of the bound on the DM density. This is illustrate in Fig.~\ref{fig1} where we have varied the DM halo density in units of the canonical value at 8 kpc, $\rho_{0} \equiv 0.3~{\rm GeV}~{\rm cm}^{-3}$. In addition we show in the blue band how the uncertainty in the timing information impacts the limit. 

We also note that our PPN bounds are somewhat stronger than the conservative estimation made in~\cite{Monitor:2017mdv} which only include the MW potential out to 100 kpc. Further, while Ref.~\cite{Wei:2017nyl} finds similar bounds on the PPN parameters, our complementary work considers the effects of the MW potential.

Finally, note that strong bounds on PPN parameters have been obtained from FRB and GRB observations~\cite{Wei:2015hwd}. These have been obtained however using the Keplerian model of the MW potential. Using the MW model outlined above, we can return to these limits. For example, the strongest bounds came from GRB 100704A which gave a Keplerian bound of, $\Delta \gamma \lesssim 4 \times 10^{-9}$. We find for the MW model however that this bound is weakened to $\Delta \gamma \lesssim 3 \times 10^{-8}$.

\section{Summary} 
We have shown that the exciting recent LIGO observation of the first NS-NS merger, GW170817 with the associated EM counterpart can provide new distinct tests on both astrophysics and fundamental physics. This demonstrates the power of multi-messenger astrophysics. 

We have discussed implications for the emission mechanism of prompt emission. SGRBs have been believed to be the gamma-ray emission from relativistic jets. If GW170817 is associated with off-axis outflows, the prompt emission radius for GRB 170817A is constrained to be $r\lesssim{10}^{12}$~cm. This radius can be consistent with the photospheric radius, but it could be jet-induced shock breakout emission. We point out that the constraints on non-GR effects can further be improved by understanding the jet formation, propagation, and resulting prompt emission mechanisms.

We have examined time-lag tests on Einstein's theory of GR. We expect these bounds to be improved with more GW and EM associations are found. In particular, we have demonstrated the relevance of more detailed modeling of the gravitational potential to place more robust limits on the gravitational time delay among different messengers. In future, not only GWs and electromagnetic waves but also neutrinos may be detected~\cite{Kimura:2017kan}, which would give us new information on the physics considered in this work.

{
\section*{Note added}
The same day our work first appeared on the arXiv, the related work~\cite{Boran:2017rdn} appeared. Our work is distinct from Ref.~\cite{Boran:2017rdn} in at least two respects: (1) In addition to WEP violations we also consider the implications of GW170817 for the gamma-ray emission site, and (2) in our calculation of the gravitationally induced time delay we consider a more detailed model of the Milky Way galaxy, while they include the impact of neighboring galaxies. We therefore consider our work complementary to Ref.~\cite{Boran:2017rdn}.}

\acknowledgements
We thank Kazumi Kashiyama and Peter M\'esz\'aros for useful discussion. The work of K. M. is partially supported by Alfred P. Sloan Foundation and the U.S. National Science Foundation (NSF) under grants NSF Grant No. PHY-1620777.

\vspace{2cm}

\bibliographystyle{JHEP}

\bibliography{nu}

\end{document}